# Neutron channeling in a nonmagnetic planar waveguide

S.V. Kozhevnikov[1]*, T. Keller[2,3], Yu.N. Khaydukov[2,3], F. Ott[4], F. Radu[5]

[1]*Frank Laboratory of Neutron Physics, JINR, 141980 Dubna Moscow Region, Russia*
*\*e-mail: kozhevn@nf.jinr.ru*
[2]*Max Planck Institut für Festkörperforschung, Heisenbergstr. 1, D-70569 Stuttgart, Germany*
[3]*Max Planck Society Outstation at FRM-II, D-85747 Garching, Germany*
[4]*Laboratoire Léon Brillouin CEA/CNRS, IRAMIS, Université Paris-Saclay, F-91191 Gif sur Yvette, France*
[5]*Helmholtz-Zentrum Berlin für Materialien und Energie, Albert-Einstein Straße 15, D-12489 Berlin, Germany*

We investigate neutron propagation in a middle layer of a planar waveguide which is a tri-layer thin film. A narrow divergent microbeam emitted from the end face of the film is registered. The neutron channeling length is experimentally measured as a function of the guiding channel width. Experimental results are compared with calculations.

PACS numbers: 03.75.Be, 68.49.-h, 68.60.-p, 78.66.-w

## I. INTRODUCTION

Neutron scattering is a powerful nondestructive method for the investigation of biological objects, polymers and magnetic structures due the particular properties of neutrons: isotopic sensitivity, strong penetration ability and intrinsic magnetic moment. Neutrons and X-rays are complementary methods because of its properties. For example polarized neutrons are used for the investigation of bulk materials what is inaccessible for X-rays with weak penetration ability.

A neutron beam size defines a spatial resolution and a scale of investigated objects. The conventional neutron beam width is from 0.1 to 10 mm. We should have very narrow neutron beams for the investigation of for the investigation of local microstructures in the scale of tenths millimeters. For this aim the various focusing devices are developed (diffraction gratings, refractive lenses, bent crystals-monochromators, etc.) [1], which are able to focus the neutron beam down to 50 μm. Less width is restricted by physical properties or treatment of used materials.

More effective focusing devices are planar waveguides transforming a conventional collimated neutron beam into a divergent microbeam compressed in one dimension to the thickness 0.1-10 μm (see Fig. 1). The incident neutron beam of the angular divergence $\delta\alpha_i$ falls to the surface of a tri-layer film under a small grazing angle $\alpha_i$, tunnels through the upper layer, propagates in the middle layer as in a channel and is emitted through the end face as a microbeam. The initial width of the microbeam is equal to the channel width $d$ and the final width depends on the microbeam angular divergence and the distance to an investigated sample. Fraunhofer diffraction $\delta\alpha_F \sim \lambda / d$ on a narrow slit which is the channel of the width $d$ mainly contributes into the microbeam divergence $\delta\alpha_f = \sqrt{\left(\delta\alpha_i\right)^2 + \left(\delta\alpha_F\right)^2}$. Therefore to keep the minimal width of the microbeam we have to place the sample close to the waveguide exit edge, collimate the incident beam, decrease the neutron wavelength and increase the channel width. If the sample is magnetic then the magnetic field on the sample should not effect to the waveguide. In such geometry the waveguide should be nonmagnetic and the incident beam should be polarized.

The theory [2] says that changing of the channel width $d$ changes the neutron wave density decay parameter termed as *the neutron channeling length*. The aim of this investigation is experimental determination of the neutron channeling length as a function of the guiding channel width and comparison of results with theoretical predictions.

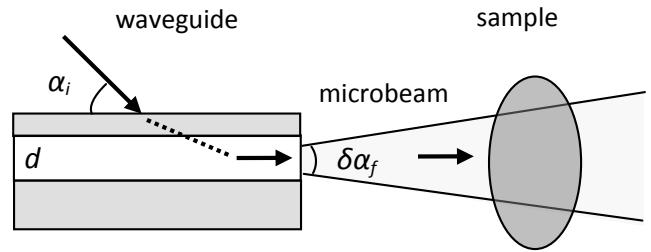

Fig. 1. Experimental geometry with neutron microbeam.

Neutron wavefunction density is resonantly enhanced inside a guiding channel of a planar waveguide. The theory of neutron resonances in layered structures is developed in [3]. There are several types of resonator structures with neutron scattering length density (SLD) in the shape of potential well.

*Interference filters* have all three layers equally thin. Its transmit neutrons in a very narrow interval of energy. Neutron transmission coefficient has one narrow maximum on energy and on the total reflection there is a corresponding narrow minimum. This phenomenon is *frustrated total reflection*. SLD of this layered structure correspond to the kinetic energy of ultracold neutrons therefore interference filters are successfully used for monochromatization and spectrometry of ultracold neutrons. In [4] the multilayer and in [5] the tri-layer interference filters were calculated. In [6,7] the first experiments with interference filters were described. Review of experiments on fundamental physics using the interference filters for ultracold neutrons can be found in [8].

For the interference filters with respectively wide middle layer there are an interference picture with many narrow minima at reflection and corresponding narrow maxima in transmission. Such structure is neutron analog of *Fabry-Perot interferometer*. In [9] there is review of experiments. It was proposed to use these interference filters for monochromatization and polarization of the neutron beam, for the measurement of a penetration depth of a magnetic field into superconducting films, etc.

If the bottom layer is relatively thick, then neutrons practically do not pass through this structure but are almost reflected from it. In the middle layer the resonantly enhanced neutron standing waves are formed. Such structure is *resonator* and used for enhancement of weak interaction of neutrons with matter. There are various ways to register neutron standing waves in layered structures: 1) minima on total reflection; 2) maxima of spin-flip neutron intensity, off-specular diffuse neutron scattering diffused, etc.; 3) maxima of secondary characteristic irradiation (gamma-rays, alpha-particles, etc.) as a result of neutron interaction with corresponding elements (Gd, $^6$Li, etc.). Detail description of production, registration, and application of neutron standing waves in planar waveguides can be found in review [10]. Incoherent neutron scattering was registered via minima on total neutron reflection [11] and directly via maxima of incoherent neutron scattering intensity [12]. In [13] minima on total neutron reflection and corresponding maxima of gamma-rays intensity from the $Gd_2O_3$ layer were registered. In [14] narrow minima on total neutron reflection and maxima of alpha-particles intensity were observed at neutron interaction with the $^6$LiF layer. In [15] it was proposed to use the resonator with the uranium layer for creation of a miniature atomic electric power station. In [16,17] neutron spin-flip intensity was registered at reflection from resonators with magnetically non-collinear layers. Off-specular diffuse neutron scattering at the interface roughness [17,18] and the interfaces near the magnetic domain structures [19-21]. In the last time the interest to application of layered resonators to investigation of magnetism is growing. In [22] a resonator was used for the investigation of coexistence of magnetism and superconductivity in films. In [12,23] resonances positions on $z$ coordinate inside a resonator were changed by polarized neutron beams. This way can be used to select desired magnetic layers for the investigation.

If layered resonators are used for neutron channeling or neutron microbeam production then such structures are called *waveguides*. For the first time a planar waveguide was considered theoretically in [24]. The principle similar to refractive lenses in an optical waveguide was proposed to introduction of the incident neutron beam into the guiding layer. Therefore such waveguides are termed as *prism-like waveguides* in contrast to *simple waveguides* in Fig. 1. The polarized neutron microbeam from the end face of the prism-like waveguide was obtained for the first time in [25]. Such waveguides have more complicated structure and therefore were not developed further. In present time the simple waveguides are used. The unpolarized [26] and polarized [27] neutron microbeams were produced using the neutron reflectometers with fixed neutron wavelength. In [28] the system of neutron microbeams was registered using the time-of-flight neutron reflectometer and Fraunhofer diffraction contribution $\delta\alpha_F \sim \lambda$ into the microbeam divergence was measured experimentally. Fraunhofer contribution $\delta\alpha_F \sim 1/d$ into the microbeam angular divergence was measured experimentally using the time-of-flight [29] and fixed wavelength [30] reflectometers. For the first time a polarized neutron microbeam from a waveguide was applied for the investigation of a microstructure in [31]. The combination of a nonmagnetic waveguide and a polarized neutron reflectometer was used [32]. Spatial scan of an amorphous magnetic wire with axial domains in a compact core and circular domains in a wide shell was done. Larmor precession of neutron spin method at transmission was used [33]. Intrinsic spectral width of the resonance was estimated experimentally in [34]. It is important for the resolution of the Larmor precession method. In review [35] various methods are compared: slits from absorbing materials, total neutron reflection from a short substrate and planar waveguides. The most versatile and high flux method is reflection from the substrate but the waveguides produce the narrowest microbeam.

The process of neutron propagation along the channel is termed as *channeling* and the parameter of the exponential decay of neutron wavefunction density is called *channeling length*. For the first time the phenomenon of neutron channeling was registered in the reflection geometry for the prism-like waveguide in [36] and for the simple waveguide in [37]. Neutrons propagate along the channel and go out not from the edge but through the upper layer in the specular reflection direction. In this case the width of the reflected beam is equal to the width of the incident primary beam. In other words, the channeled and then reflected beam is not a microbeam by its width but has a resonance nature. In [38] it is proposed to use polarized neutron channeling (PNC) for the direct measurement of magnetization of thin films with high precision. But calculations were done for the prism-like waveguide. In [39] the calculations were done for the simple waveguide with a weakly magnetic film as a middle layer. In [40,41] using PNC the weakly magnetic films containing rare-earth elements were investigated. Such films are broadly used for the development of new methods of magnetic recording and switching [42]. The polarized neutron microbeam intensity from the end face is registered vs. the grazing angle of the incident beam. Difference of the peaks positions for spin up and down allows to directly extract magnetization of the film with precision around 10 G. In [43-46] this and other direct neutron methods for the investigation of magnetic films are discussed: Larmor precession, spatial beam-splitting and neutron spin resonance in matter.

The theory of neutron channeling in layered structures was developed in [2]. According to this theory, the neutron channeling length depends on the parameters of the waveguide (upper layer thickness, channel width, potential well depth) and the resonance order. For the first time the neutron channeling length was experimentally measured in [47,48]. The experimental setup and the ways to the channeling length measurement are described in [49]. Recently we obtained experimentally following. The channeling length grows exponentially with the increasing of the upper layer thickness [49,50] and linear with increasing of SLD of outer layers at the fixed SLD of the channel [51] and decreases inversely with increasing of the resonance order [50]. Experimental results proved the theoretical predictions [2]. Then we will consider the dependence of the channeling length on the channel width.

## II. CALCULATIONS

The planar neutron waveguide is tri-layer structure $Ni_{67}Cu_{33}(20\ nm)/Cu(d)/Ni_{67}Cu_{33}(50)//Al_2O_3$ (substrate) having SLD in the shape of potential well (Fig 2a). The material Ni(67 ат. %)Cu(33 ат. %) is nonmagnetic at room temperature. The upper thin layer $Ni_{67}Cu_{33}$ and the bottom thick layer $Ni_{67}Cu_{33}$ have the high SLD and the middle thick layer has the low SLD. Neutron beam in vacuum (medium 0) falls on the waveguide surface under the small grazing angle $\alpha_i$ (Fig. 2b). Then neutrons tunnel through the upper thin layer $Ni_{67}Cu_{33}$ of the thickness $a$ (medium 1), enter to the middle layer Cu of the thickness $d$ (medium 2) and are almost totally reflected from the bottom thick layer $Ni_{67}Cu_{33}$ (medium 3). Then the part of neutrons tunnels from the middle layer through the upper layer and goes out from the surface. Another part of neutrons propagates along the layers in the $x$ direction and is emitted from the edge as the divergent microbeam. The neutron wavefunction amplitude is enhanced at the resonance conditions for the phase of the neutron wavefunction inside the guiding layer [3]:

$$\gamma(k_{0z}) = 2k_{2z}d + \arg(R_{21}) + \arg(R_{23}) = 2\pi n \quad (1)$$

where $k_{0z} = k_0 \sin \alpha_i$ is $z$ - projection of the neutron wave vector in vacuum; $k_{2z} = \sqrt{k_{0z}^2 - \rho_2}$ is $z$ - projection of the neutron wave vector in the guiding layer; $\rho_2$ is SLD of the guiding layer; $R_{21}$ is the neutron reflection amplitude from the upper layer at propagation in the guiding layer; $R_{23}$ is the neutron reflection amplitude from the bottom layer at propagation in the guiding layer; n=0, 1, 2 ... is the resonance order.

The part of neutrons leaks from the guiding layer through the upper layer. Therefore the neutron wavefunction density is exponentially decays $\sim \exp(-x/x_e)$ on the characteristic distance $x_e$ called the neutron channeling length. In the theory of neutron channeling [2] the channeling length can be found:

$$x_e \approx k_x d / k_{2z} T \qquad (2)$$

where $k_x = k_0 \cos \alpha_i$ is $x$ - projection of the neutron wave vector; $T$ is the neutron transmission coefficient through the upper layer to vacuum which depends on the upper layer thickness as $T \sim \exp(-2k_{1z}a)$ where $k_{1z} = \sqrt{\rho_1 - k_{0z}^2}$ is $z$ - projection of the neutron wave vector in the upper layer; $\rho_1$ is SLD of the upper layer. From (1) it follows that the energy of the resonance depends directly on the resonance order, the channel width and SLD of the channel. By the indirect way via the reflection amplitudes $R_{21}$ and $R_{23}$ the energy of the resonance depends on SLD and the thickness of the upper layer. And the neutron channeling length depends on the energy of the resonance. It means that it is impossible to define exact analytical function of the neutron channeling dependence on one parameter of the waveguide. It is necessary to carry out numerical calculations using (2) and (1). The channeling length is measured experimentally for the real parameters of the waveguide (SLD and thicknesses of layers). Neutron reflectometry defines the fitted parameters only with the accuracy depending on several factors (statistical errors bar, resolution function of experimental setup, model of calculations, quality of fabricated structure, etc.). Usually the fitted parameters are close to real parameters with small corrections. Therefore this small difference should not change the function of the channeling length dependence (linear, exponential, reversal). It means that we may use the qualitative comparison of experimental results with calculations.

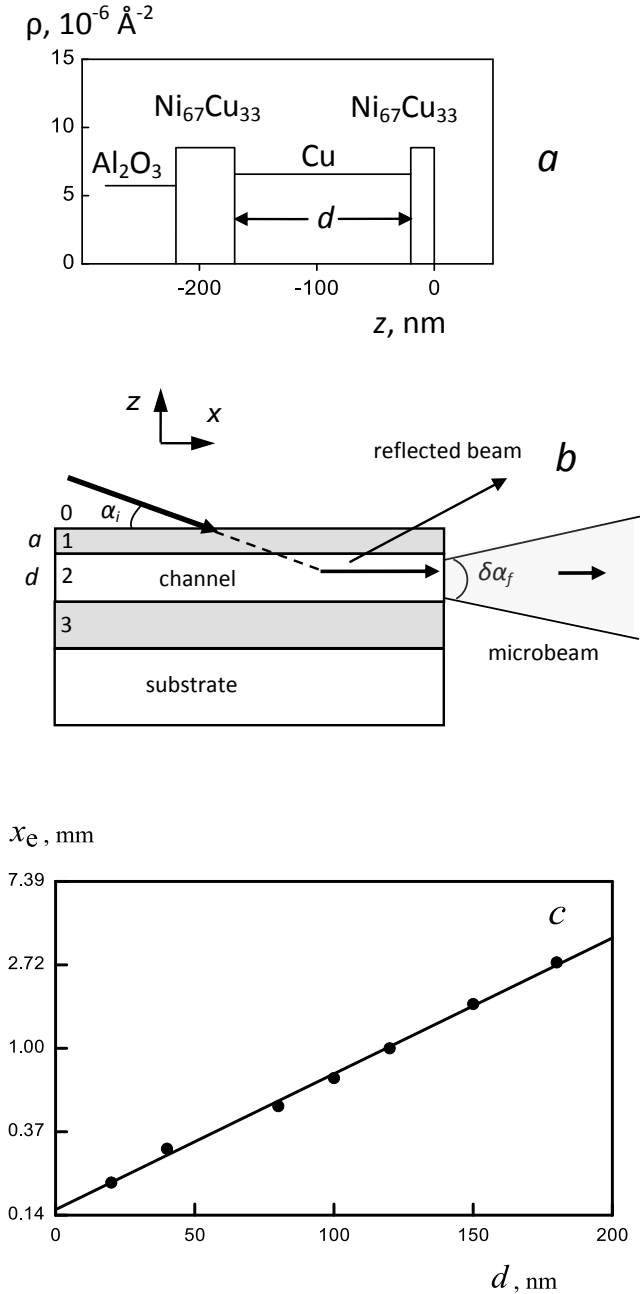

Fig. 2. On calculations of the neutron channeling length in planar waveguides: (*a*) SLD of the planar waveguides vs. the coordinate $z$ perpendicular to the layers; (*b*) geometry of neutron channeling in planar waveguides; (*c*) neutron channeling length calculated vs. the channel width.

The calculated by (2) and (1) neutron channeling length for the resonance n=0 vs. the channel width is shown in In Fig. 2*c*. Used SLD are from the table (see Fig. 2*a*). The dependence is linear in the natural logarithm scale:

$$\ln x_e \sim d \qquad (3)$$

Thus the theory predicts that the neutron channeling length for the resonance n=0 should exponentially increase with increasing the channel width.

### III. EXPERIMENT

Experiment was done using the neutron reflectometer NREX with the horizontal sample plane at the steady state reactor FRM II (MLZ research center, Garching, Germany). The fixed neutron wavelength is 4.26 Å (resolution 1 % FWHM). Nutron beam was registered by the gas $^3$He two-coordinate position-sensitive detector with the spatial resolution 3 mm. The angular divergence of the incident beam 0.0065° is defined by the first slit of the width 0.25 mm placed after the monochromator. The distance from the first slit to the sample was 2200 mm and from the sample to the detector was 2400 mm. The second slit of the width 0.7 mm was placed at 200 mm before the sample was used to reduce background.

Using the unpolarized neutron beam the set of four nonmagnetic samples with the nominal structure Ni$_{67}$Cu$_{33}$(20 нм)/Cu($d$)/Ni$_{67}$Cu$_{33}$(50 нм)//Al$_2$O$_3$ (substrate) were investigated where $d$ = 80, 100, 120, 180 (nm). The substrate sizes were 10×10×0.5 mm$^3$ but for the sample of $d$ = 100 nm the substrate had the sizes 10(along beam)×20×1 mm$^3$. The fifth sample Ni$_{67}$Cu$_{33}$(20 нм)/Cu(150)/Ni$_{67}$Cu$_{33}$(50)//Si(substrate) with the substrate sizes 25×25×1 mm$^3$ was investigated in [50].

Two-dimensional map of neutron intensity for the sample with the channel width 180 nm is shown in Fig. 3 vs. the grazing angles of the incident and scattered beams. The horizontal dashed line $\alpha_f = 0$ corresponds to the sample plane. In the bottom the high intensity refracted beam as a diagonal is seen. Ovals mark the microbeam spots of the resonance orders n=0, 1, 2, 3, 4. The microbeam intensity shown symmetrical respect to the sample plane direction but the bottom part of the microbeams is covered by more intensive refracted beam.

In Fig. 4 the neutron microbeam intensity is presented vs. На рис. 4 представлена интенсивность микропучка в зависимости от угла скольжения начального пучка $\alpha_i$. The intensity is summarized on the final angle $\alpha_f$ in the upper interval in Fig. 3 where the microbeam is not covered by the refracted beam. For the channel width 180 nm (Fig. 4$a$) the resonance peaks are close to each other and partially overlapped because of the wide channel thickness. From the resonance condition (1) it follows that the distance between resonances decreases with the channel width increasing. In Fig. 4$b$ the resonances for the sample with the channel width 120 nm are resolved badly. It may be caused of the thin substrate bending or the bad quality of the sample. The highest microbeam intensity with good resolved peaks is in Fig. 4$c$. The reasons are following: 20 mm length for the incident beam (other samples have 10 mm), the thick non-bending substrate and the small width of the channel. The sample with the smallest channel width 80 mm has the largest distance between peaks but the ratio *signal/background* is low because of the big divergence of the macrobeam (Fig. 4$d$).

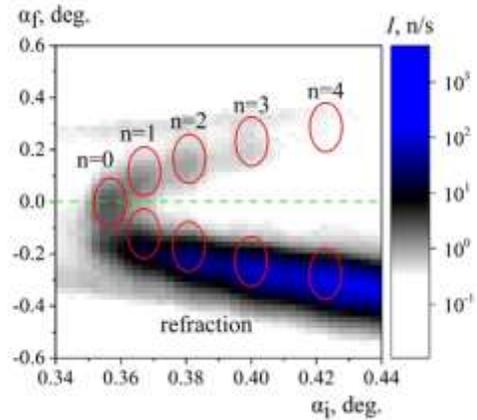

Fig. 3. Neutron intensity vs. the initial and final grazing angles for the channel width 180 nm. Ovals mark the neutron microbeams of the resonances n=0, 1, 2, 3, 4.

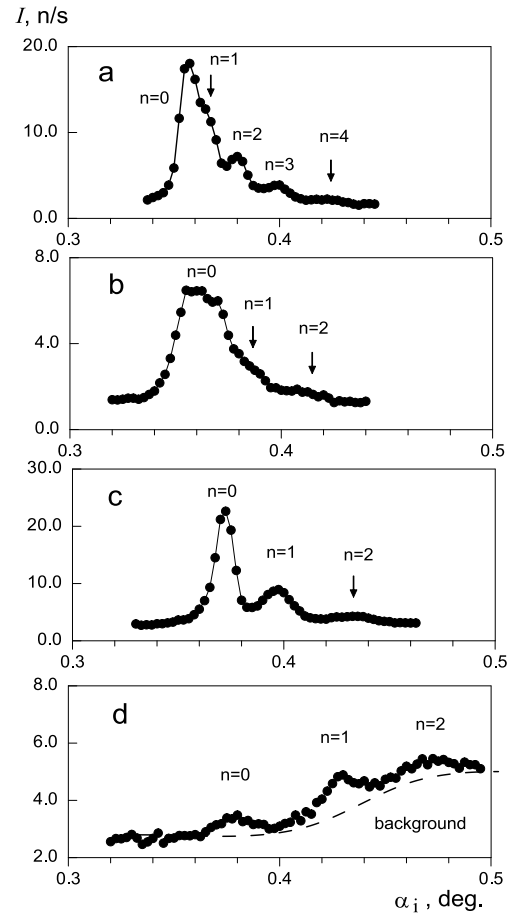

Fig. 4. The neutron microbeam intensity vs. the grazing angle of the incident beam for the different channel width: ($a$) 180 nm; ($b$) 120 nm; ($c$) 100 nm; ($d$) 80 nm.

In Fig.5 the neutron microbeam intensity vs. the final angle $\alpha_f$ is presented for the resonance n=0 and the various channel width (*a-d*) and for the channel width 180 nm and the resonance orders n=1, 2, 3, 4 (*e-h*). Symbols and line corresponds to experiment and calculations respectively. In [26,50] the experimental angular distribution of the microbeam intensity is described by Fourier transformation of the neutron wavefunction inside the waveguide. In Fig. 5 the calculations also describe the experimental data.

The resonance n=0 has the intensive central peak and the resonances of higher orders have the outer peaks of high intensity and low intensity between them. One can see that the left side of the microbeam is covered by the refracted beam. The microbeam divergence of the resonance n=0 increases with decreasing the channel width. The microbeam with the channel width 80 nm has the big angular divergence and low intensity and therefore is overlapped by the refracted and reflected beams (Fig. 5*d*).

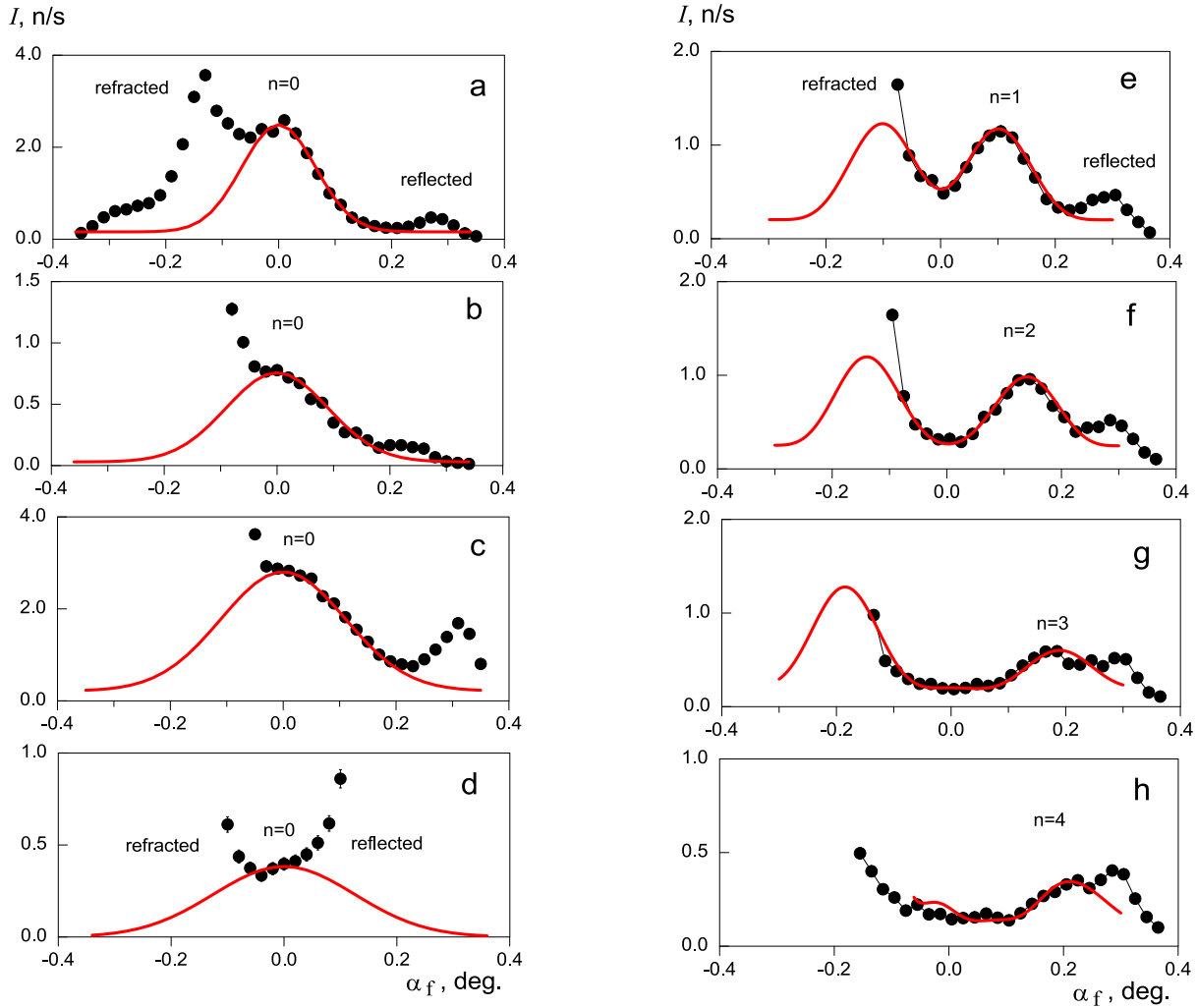

Fig. 5. Neutron microbeam intensity vs. the final angle for the resonance n=0 and the different channel width: (*a*) n=1; (*f*) n=2; (*g*) n=3; (*h*) n=4. Points and thick lines correspond to the experimental data and the calculations using Fourier transformation of the neutron wavefunction density in the waveguide, respectively.

We used the material boral (aluminum with absorbing neutrons boron) for the measurement of the neutron channeling length. The absorber bar was placed on the waveguide near the exit end face (Fig. 6$a$). The bar cross-sections were $1{\times}1$ mm$^2$ and $2{\times}2$ mm$^2$ and its lengths were 38 and 30 mm. More detailed description of the experimental setup and the method was done in [49]. Using the micrometric translation table the bar is moved on the surface from the exit edge and covers the part of the surface from the illumination by the incident neutron beam. There is an air gap of the height $h$ under the bar due to the bar curvature. The real non-illuminated length of the surface $x$ is less than the distance $L$ from the exit edge to the front edge of the bar: $x = L - \Delta x$. The value $\Delta x$ depends on the gap height, the grazing angle of the incident beam and imperfection of the sample installation respect to the exit edge of the sample. For this method we do not need to know the value $\Delta x$. It is different for each experiment. But for reference it was defined in [51] and consisted of around 10 μm what corresponded to $\Delta x$ around 1.5 mm.

The results obtained in [50] for the waveguide structure Ni$_{67}$Cu$_{33}$(20 nm)/Cu(150)/Ni$_{67}$Cu$_{33}$(50)//Si(substrate) are shown in Fig. 6$b$ for explatation of this method. We measure the microbeam intensity $I$ ($x = 0$) at the fully illuminated sample surface. Then the bar is translated and the microbeam intensity is registered vs. the distance $L$ from the exit edge of the sample and the front edge of the absorber. Then we plot the normalized microbeam intensity $I$ ($L$) / $I$ ($x = 0$) in the natural logarithm scale (the upper $L$ axis of abscises). All point lay on the direct line crossing the level of 1 on the ordinates axis in a point. This point corresponds to the moment when the absorber begins to cover the sample surface from the illumination by the incident neutron beam. This point defines the correction value $\Delta x$ to shift all experimental points on $L$ dependence. We obtain the experimental dependence of the normalized microbeam intensity on the length $x$ of the non-illuminated sample surface (bottom axis of abscises). The point of the normalized intensity decreasing in e times corresponds to the measured channeling length $\Delta x = 1.7 \pm 0.2$ (mm).

The neutron microbeam intensity of the resonance n=0 vs. the grazing angle of the incident beam for the different distance $L$ is shown in Figs. 7-10. The samples are Ni$_{67}$Cu$_{33}$(20 nm)/Cu($d$)/Ni$_{67}$Cu$_{33}$(50 nm)//Al$_2$O$_3$ (substrate) with the different channel width: 80 nm (Fig. 7); 100 nm (Fig. 8); 120 nm (Fig. 9) and 180 nm (Fig. 10). The microbeam intensity was integrated in the narrow interval of the final angles $\alpha_f$ around the sample plane direction where the microbeam is not covered by the refracted beam. The dashed line indicates a background level. For the channel width 180 nm (Fig. 10$a$) the part of the microbeam intensity of the resonance n=1 was extracted as the parasitic background.

The normalized microbeam intensity in the natural logarithm scale vs. the length of non-illuminated part of the sample surface $x$ is shown in Fig. 11 for the different channel width: ($a$) 180 nm; ($b$) 120 nm; ($c$) 100 nm; ($d$) 80 nm. The microbeam intensity is integrated under the peaks in Figs. 7-10 and normalized on the microbeam intensity at the fully illuminated surface. Then the correction on $\Delta x$ due to the air gap was done. From the lines in Fig. 11 the experimental channeling length was defined.

The neutron channeling length in the scale of the natural logarithm vs. the channel width is presented in Fig. 12. The error bar is defined by the extreme trajectories passing via error bars in Fig. 11. In [29] the parameters of the samples

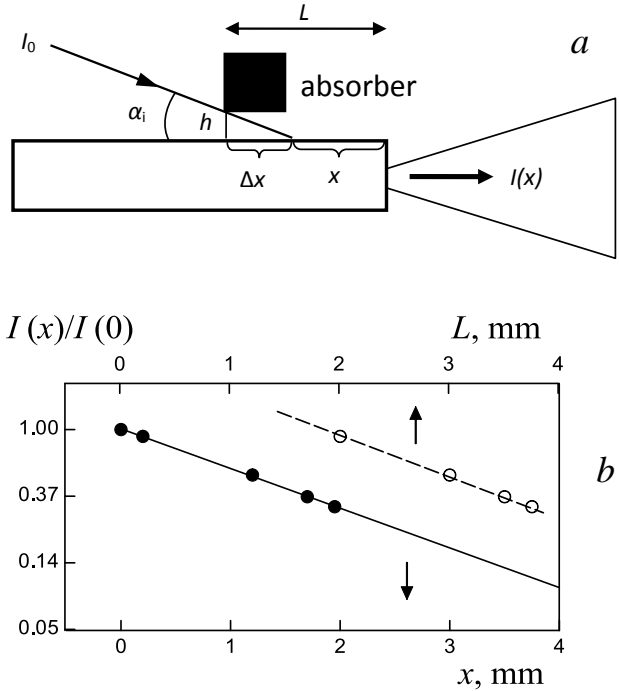

Fig. 6. The measurement of the channeling length: ($a$) geometry of experiment; ($b$) the neutron microbeam normalized intensity vs. the distance $L$ (upper axis) and vs. the length $x$ of the non-illuminated part of the sample surface (lower axis) [50]. The channel width is 150 nm. Open and closed symbols correspond to raw and corrected on $\Delta x$ data, respectively.

Ni$_{67}$Cu$_{33}$(20 nm)/Cu($d$)/Ni$_{67}$Cu$_{33}$(50 nm)//Al$_2$O$_3$ (substrate) were defined using fit of neutron reflectivities. The channel widths are close to the nominal values. Points in Fig. 12 are experimental results and line is the approximation by exponential function. One can see that the obtained experimental data are satisfactory described by the straight line and qualitative corresponds to the theoretical predictions in Fig. 2$c$. Thus we have experimentally proved that the channeling length is exponentially increased with increasing the channel width.

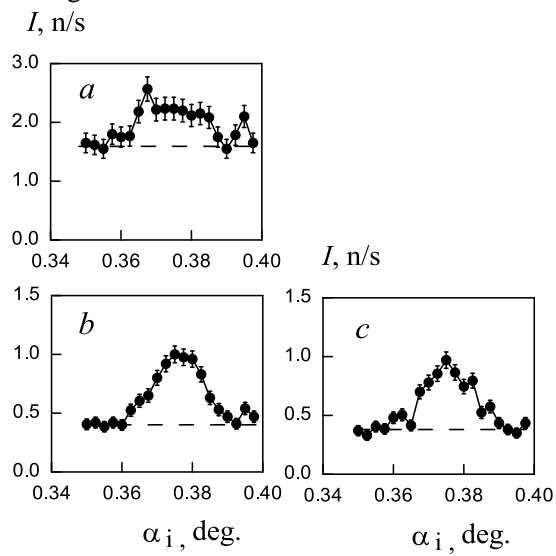

Fig. 7. Neutron microbeam intensity vs. the grazing angle of the incident beam $\alpha_i$ for the channel width 80 nm and different distance $L$ from the exit edge of the waveguide to the front edge of the absorber: ($a$) without absorber; ($b$) 2.2 mm; ($c$) 2.4 mm.

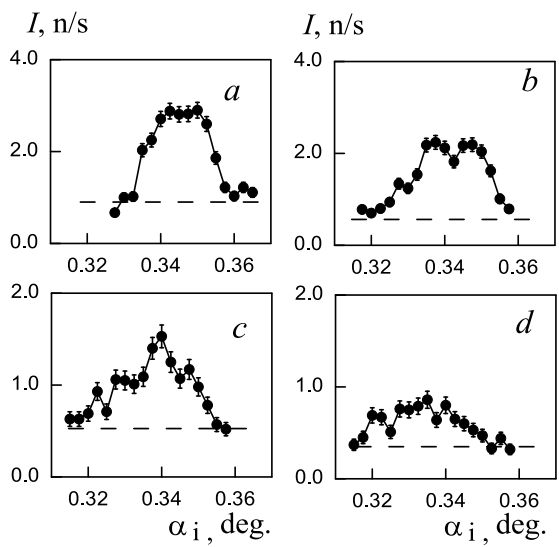

Fig. 9. Neutron microbeam intensity vs. the initial angle $\alpha_i$ for the channel width 120 nm and different distance $L$: ($a$) without absorber; ($b$) 1.7 mm; ($c$) 2.0 mm; ($d$) 2.3 mm.

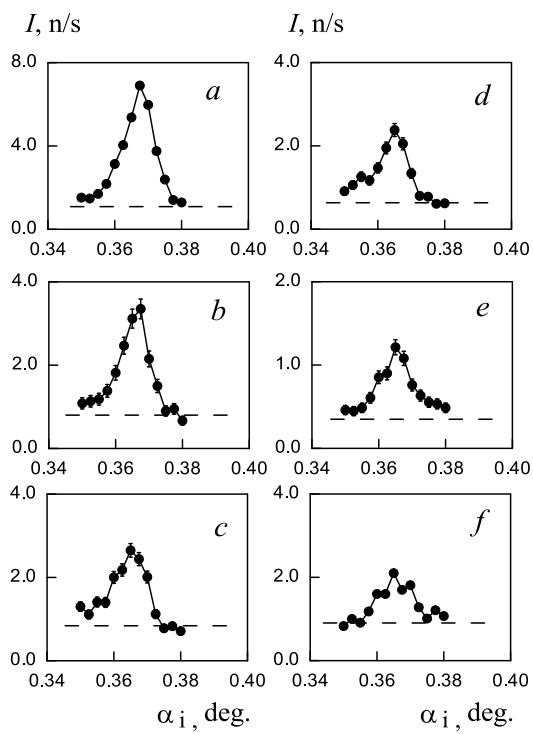

Fig. 8. Neutron microbeam intensity vs. the initial angle $\alpha_i$ for the channel width 100 nm and different distance $L$: ($a$) without absorber; ($b$) 2.0 mm; ($c$) 2.2 mm; ($d$) 2.4 mm; ($e$) 2.7 mm; ($f$) 3.2 mm.

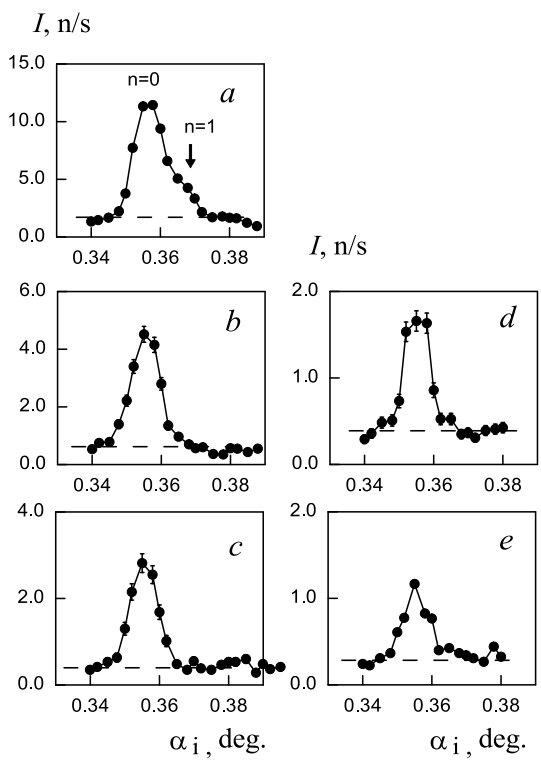

Fig. 10. Neutron microbeam intensity vs. the initial angle $\alpha_i$ for the channel width 180 nm and different distance $L$: ($a$) without absorber; ($b$) 4.0 mm; ($c$) 4.5 mm; ($d$) 5.0 mm; ($e$) 5.5 mm.

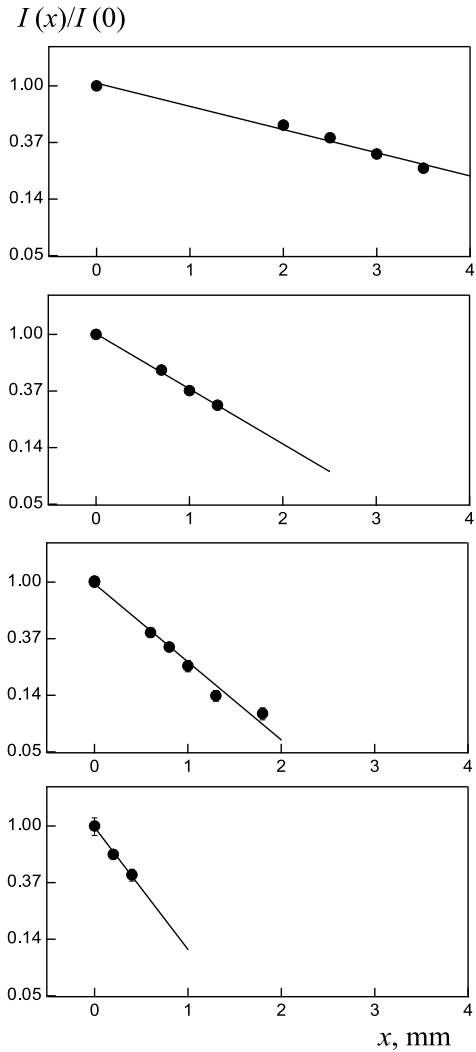

$I(x)/I(0)$

$x$, mm

Fig. 11. The neutron microbeam normalized intensity vs. the length $x$ of the non-illuminated part of the sample surface for the different channel width: (*a*) 180 nm; (*b*) 120 nm; (*c*) 100 nm; (*d*) 80 nm.

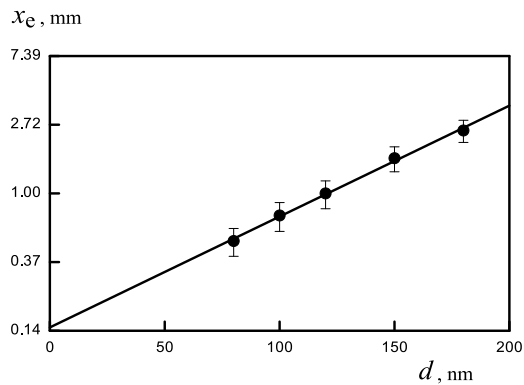

$x_e$, mm

$d$, nm

Fig. 12. The experimentally measured neutron channeling length in the natural logarithm scale vs. the channel width. Line is the fit by exponential function.

## IV. CONCLUSIONS

The parameter of exponential decay of the neutron wavefunction density in the tri-layer planar waveguide was defined experimentally. The absorbing bar placed on the sample surface was used. We measured the neutron microbeam intensity emitted from the end face of the channel vs. the length of the non-illuminated sample surface under the absorber. It was obtained that the channeling length was exponentially increased with the increasing of the guiding channel width. This result confirms qualitatively the prediction of the theory of the neutron channeling in planar waveguides.

Earlier we used a combination of a non-magnetic waveguide with a polarized neutron reflectometer for the high-resolution spatial scan of a bulk magnetic microstructure using a polarized neutron microbeam. The spatial resolution of this method depends on the microbeam divergence which decreases with increasing the channel width. Thus we can control the produced microbeam properties by changing the parameters of the waveguide. And the properties of the neutron wave function inside the waveguide are also changed. The problem how the microbeam intensity depends on the waveguide parameters is not solved yet. We hope that the results obtained in this work will be useful in the future for the neutron microbeam intensity optimization and planning of experiments.

### Acknowledgements

The authors thank V.K. Ignatovich for useful discussions, V.L. Aksenov and Yu.V. Nikitenko for the interest to this subject. The work is supported by the JINR-Romania scientific project No. 323/21.05.2018, items 89 and 90.